\documentclass[a4paper,11pt]{article}
\usepackage{graphicx,epsfig,picins}
\usepackage{amsfonts,amsmath,amssymb}
\usepackage{fancyhdr,fancybox}
\usepackage{indentfirst}
\usepackage{titlesec}
\usepackage{verbatim}
\usepackage[sort&compress, numbers]{natbib}
\usepackage{array,dcolumn,tabularx}
\usepackage{bm}
\usepackage{setspace}

\begin{document}

%图标题
\renewcommand{\figurename}{Figure}
%表格标题
\renewcommand{\tablename}{Table}
%参考文献标题
\renewcommand{\refname}{References}

%在这里填写页眉和页脚
%\pagestyle{fancy} \lhead{} \chead{} \rhead{} \lfoot{} \cfoot{\thepage} \rfoot{}
%\renewcommand{\headrulewidth}{0.4pt}

%在这里填写标题和作者
\title{\textbf{Kinetic behavior of the general modifier mechanism of Botts and Morales with non-equilibrium binding}}
\author{Chen Jia$^{1}$,\;\;\;Xu-Feng Liu$^{1}$,\;\;\;Min-Ping Qian$^{1,2}$,\;\;\;Da-Quan Jiang$^{1,*}$,\;\;\;Yu-Ping Zhang$^{1,2}$}
\date{}
\maketitle

\thispagestyle{fancy} \fancyhf{}
\renewcommand \headrulewidth{0pt}
\renewcommand \footrulewidth{0.7pt}
\fancyfoot[LF]{\scriptsize
$^1$School of Mathematical Sciences, Peking University, Beijing 100871, China.\\
$^2$Center for Theoretical Biology, Peking University, Beijing 100871, China.\\
$^*$Corresponding author; Tel: 86-10-62755615; Fax: 86-10-62751801;
Email: jiangdq@math.pku.edu.cn}
\fancyfoot[CF]{\ \notag\\
\ \notag\\1}

% 在这里填写摘要
\begin{abstract}
In this paper, we perform a complete analysis of the kinetic behavior of the general modifier mechanism of Botts and Morales in both equilibrium steady states and non-equilibrium steady states (NESS). Enlightened by the non-equilibrium theory of Markov chains, we introduce the net flux into discussion and acquire an expression of product rate in NESS, which has clear biophysical significance. Up till now, it is a general belief that being an activator or an inhibitor is an intrinsic property of the modifier. However, we reveal that this traditional point of view is based on the equilibrium assumption. A modifier may no longer be an overall activator or inhibitor when the reaction system is not in equilibrium. Based on the regulation of enzyme activity by the modifier concentration, we classify the kinetic behavior of the modifier into three categories, which are named hyperbolic behavior, bell-shaped behavior, and switching behavior, respectively. We show that the switching phenomenon, in which a modifier may convert between an activator and an inhibitor when the modifier concentration varies, occurs only in NESS. Effects of drugs on the Pgp ATPase activity, where drugs may convert from activators to inhibitors with the increase of the drug concentration, are taken as a typical example to demonstrate the occurrence of the switching phenomenon.
\end{abstract}

%在这里填写关键字

%开始正文
\section{Introduction}
% Modifiers
Modifiers or effectors, ligands that bind to enzymes and thereby alter their catalytic activity, play a crucial role in the study of biochemical problems, e.g., enzymatic catalysis and metabolic pathways \cite{Cornish-Bowden, Todhunter, Bertucci, Malykh, Conway}. Moreover, they have wide applications in pharmacology, toxicology, industry and agriculture. Activators and inhibitors are defined as modifiers that strengthen or weaken, respectively, the enzyme activity of the reaction system \cite{Segel, Fontes}. The enzyme activity is generally characterized in terms of the rate of product formation of the enzyme-catalyzed reaction in the steady state.

\begin{figure}[ht]
\begin{center}
\centerline{\includegraphics[width=.5\textwidth]{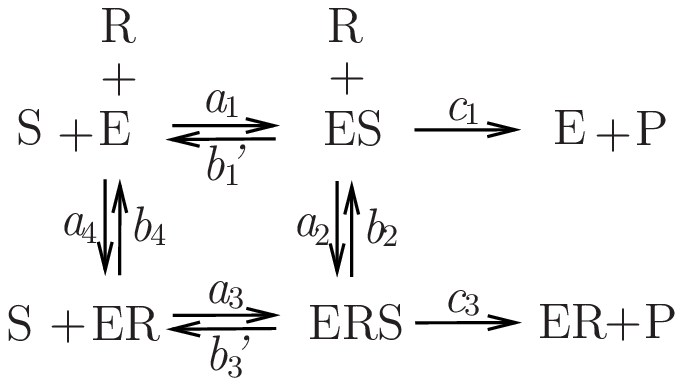}}
\caption{}\label{mechanism}
\end{center}
\end{figure}

% General modifier mechanism of Botts and Morales
Most enzyme mechanisms that involve a modifier reversibly acting on Michaelis-type enzymes can be regarded as a particular case of the general modifier mechanism of Botts and Morales, as is depicted in Fig.~\ref{mechanism} \cite{Botts}. Many theoretical biologists have studied the steady state and transient phase kinetics of the general modifier mechanism \cite{Segel, Fontes, Botts, Segel-Martin, Topham, Schmitz, Topham-Brocklehurst, DiCera, Varon1999, Varon2002, Al-Shawi2003} and its particular cases, in which modifiers act on Michaelis-type enzymes as competitive inhibitors, uncompetitive inhibitors or pure non-competitive inhibitors \cite{Laidler, Cornish-Bowden, Varon2001a, Varon2001b}.

% Early works on classification of the modifier
Segel and Martin \cite{Segel, Segel-Martin} reported a steady state rate equation that is second degree in both substrate concentration $[S]$ and modifier concentration $[R]$. They also found several conditions under which the rate equation can be reduced to one that is first degree in $[S]$ and in $[R]$. Fontes et al. \cite{Fontes} discussed the behavior of the modifier with the change of substrate concentration $[S]$ under the assumption of rapid equilibrium. Laidler \cite{Laidler} studied the behavior of the modifier with the change of modifier concentration $[R]$ under some simplifying assumptions. He also suggested definitions of competitive, uncompetitive and noncompetitive activation, by analogy with the generally accepted definitions for inhibition.

% Flaws of early works
The major differences among the contributions of these authors are the set of simplifying assumptions made about the steady state reached by the enzyme-catalyzed reaction system \cite{Varon2005}. However, to date, there is still a lack of a complete analysis about the steady state kinetics of the general modifier mechanism of Botts and Morales without any simplifying assumptions. The major difficulty lies in the fact that the traditional approaches in solving this problem obscure the essence, which can hardly be realized without the concept of the net flux introduced in this article, of the enzyme system to some extent.

% NESS in the general modifier mechanism
It is illuminating to point out that the modifier- and substrate-binding steps are not dead-end reactions in the enzyme system, and so they are not necessarily in equilibrium \cite{Cornish-Bowden}. We have good reasons to believe that biochemical systems in living cells generally operate in a state far from equilibrium. Whether the cyclic reaction mechanism in Fig.\ref{mechanism} satisfies the equilibrium assumption, i.e. detailed balance, depends on whether the system is closed or open. A closed system will finally approach an equilibrium steady state, whereas an open system, driven by an external source of energy, tend to reach a non-equilibrium steady state (NESS) \cite{Hill, Wyman, Qian2007, Qian2008}.

% Circulation theory of the Markov chain
In this article, we remove the equilibrium assumption and provide a general analysis of the kinetic behavior of the modifier in NESS. Enlightened by the circulation theory of Markov chains \cite{Jiang}, we introduce the net flux into discussion and acquire an expression of product rate in NESS, which has clear biophysical significance. The essence of the general modifier kinetics is then revealed to be the competition between the equilibrium and non-equilibrium effects.

% Our contributions
So far, it is a general belief that a modifier acts as either an activator or an inhibitor for all its possible concentration values $[R]$ when the substrate concentration $[S]$ is fixed. However, we find that a modifier cannot be regarded as an overall activator or inhibitor when the reaction system is in NESS. According to our results, a particular modifier may convert from an activator to an inhibitor or vice versa with the change of $[R]$. More specifically, we classify the kinetic behavior of the modifier into three categories, which are named hyperbolic behavior, bell-shaped behavior, and switching behavior, respectively. The latter two kinds of behavior will never occur in an equilibrium steady state.

% example of drug
Incidentally, drugs are typical modifiers in pharmacology. The presence of the drug can activate or inhibit the enzyme activity. Experimental data show that a drug can always act as an activator regardless of its concentration, or first act as an activator then, from a certain concentration value, transit to be an inhibitor \cite{Al-Shawi2003}. Here the occurrence of switching phenomenon provides strong support for the argument presented in this paper.

\section{Methods}

\subsection{Catalytic cycle}
\begin{figure}[ht]
\begin{center}
\centerline{\includegraphics[width=.4\textwidth]{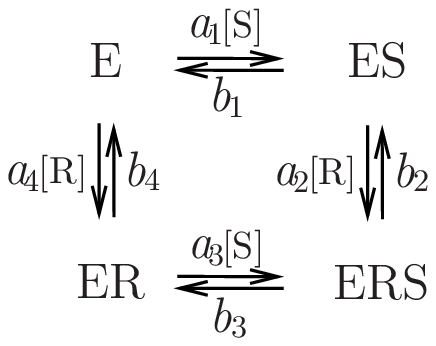}}
\caption{}\label{cycle}
\end{center}
\end{figure}

In this article, the symbols $E,\;S,\;R$ and $P$ stand for the enzyme molecule, the substrate molecule, the modifier molecule, and the product molecule, respectively, while the composite symbols $ES,\;ER$ and $ERS$ represent the corresponding complexes. If there is only one enzyme molecule, it may convert among four states: the free (unbound) enzyme $E$, the complex $ES$, the complex $ERS$ and the complex $ER$. Then from the perspective of the single enzyme molecule, the kinetics are stochastic and cyclic, as is shown in Fig.~\ref{cycle}, with pseudo-first-order rate constants $a_1[S],\;a_2[R],\;a_3[S]$, and $a_4[R]$, and first-order rate constants $b_1 = b_1'+c_1,\;b_2,\;b_3 = b_3'+c_3$, and $b_4$. We add the rate constants $b_1'$ and $c_1$, since there are two ways of transition from the complex $ES$ to the free enzyme $E$. The rate constants $b_3'$ and $c_3$ are added for the same reason.

\subsection{Net flux}
Based on the law of mass action, we have the following kinetics equations:
\begin{equation}\label{mass action}\left\{
\begin{split}
\frac{d[E]}{dt} &= -(a_1[S]+a_4[R])[E] + b_1[ES] + b_4[ER], \\
\frac{d[ES]}{dt} &= a_1[S][E] -(b_1+a_2[R])[ES] + b_2[ERS], \\
\frac{d[ERS]}{dt} &= a_2[R][ES] -(b_2+b_3)[ERS] + a_3[S][ER], \\
\frac{d[E]}{dt} &= a_4[R][E] + b_3[ERS] -(b_4+a_3[S])[ER],
\end{split}\right.
\end{equation}
where $b_1=b_1'+c_1$, $b_3=b_3'+c_3$. The above four equations constitute a system of linear equations with coefficient matrix
\begin{equation}\label{Q_matrix}
\begin{split}
Q=\left(
  \begin{array}{cccc}
    -(a_1[S]+a_4[R]) & b_1 & 0 & b_4 \\
    a_1[S] & -(b_1+a_2[R]) & b_2 & 0 \\
    0& a_2[R] & -(b_2+b_3) & a_3[S] \\
    a_4[R] & 0 & b_3 & -(b_4+a_3[S]) \\
  \end{array}
\right).
\end{split}
\end{equation}
Let $E_0 = [E]+[ES]+[ER]+[ERS]$ be the total enzyme concentration. Then the quantities $\mu_{E}=[E]/E_0,\;\mu_{ES}=[ES]/E_0, \;\mu_{ERS}=[ERS]/E_0$, and $\mu_{ER}=[ER]/E_0$ represent the probability distribution of the four enzyme states, respectively. The whole setup is nothing but a continuous-time Markov chain with transition density matrix $Q^t$ (the transpose of the matrix $Q$).

It should be indicated that the enzyme-modifier and enzyme-substrate interactions often involve rapid binding steps followed by a slow conformational change or chemical step \cite{Wang}. Thus, the quasi-steady approximation can be applied based on the difference in timescales between the catalytic cycle kinetics and the overall rate of change of biochemical reactions \cite{Qianbook}. Assuming that the cycle kinetics represented in Fig.~\ref{cycle} are rapid and maintain the enzyme and the complexes in a rapid quasi-steady state, we can obtain the steady state rate, $v$, of product formation for the general modifier kinetics:
\begin{equation}
v = \frac{d[P]}{dt} = c_1[ES]+c_3[ERS] = E_0(c_1\mu_{ES}+c_3\mu_{ERS}).
\end{equation}
For simplicity, we normalize the product rate $v$ and write
\begin{equation}\label{velocity}
v = c_1\mu_{ES}+c_3\mu_{ERS}
\end{equation}
hereinafter.

In order to obtain the steady state concentrations $[E]$, $[ES]$, $[ER]$, and $[ERS]$, we set $d[E]/dt = d[ES]/dt = d[ERS]/dt = d[ER]/dt = 0$. Then Eq.~\eqref{mass action} reduces to the following compact form:
\begin{equation}\label{muQ}
Q\mu  = 0,
\end{equation}
where $\mu = (\mu_{E},\mu_{ES},\mu_{ERS},\mu_{ER})^t$ is a column vector representing the steady state probability distribution of the four enzyme states.

In the spirit of the circulation theory of Markov chains \cite{Jiang}, we introduce the net fluxes, namely the differences between fluxes in clockwise direction and fluxes in counter-clockwise direction along adjacent states of the catalytic cycle depicted in Fig.~\ref{cycle}. When the reaction system reaches a steady state, the net fluxes between all adjacent states along the catalytic cycle are identical. Denote by $J$ the net flux along the cycle $E\rightarrow ES\rightarrow ERS\rightarrow ER\rightarrow E$. With the net flux $J$ introduced above, Eq.~\eqref{muQ} can be simplified to the following circulation equations:
\begin{equation}\label{circulation}\left\{
\begin{split}
J &= a_1[S]\mu_{E} - b_1\mu_{ES}, \\
J &= a_2[R]\mu_{ES} - b_2\mu_{ERS}, \\
J &= b_3\mu_{ERS} - a_3[S]\mu_{ER}, \\
J &= b_4\mu_{ER} - a_4[R]\mu_{E}.
\end{split}\right.
\end{equation}

\subsection{Equilibrium and non-equilibrium steady states}
From the perspective of a single enzyme molecule, the enzyme system depicted in Fig.~\ref{mechanism} is in detailed balance if the net flux $J$ along the catalytic cycle depicted in Fig.~\ref{cycle} is zero, i.e.
\begin{equation}\label{DB}\left\{
\begin{split}
a_1[S]\mu_{E} &= b_1\mu_{ES}, \\
a_2[R]\mu_{ES} &= b_2\mu_{ERS}, \\
b_3\mu_{ERS} &= a_3[S]\mu_{ER}, \\
b_4\mu_{ER} &= a_4[R]\mu_{E}.
\end{split}\right.
\end{equation}
In this article, a steady state that satisfy the condition of detailed balance described in Eq.\eqref{DB} is referred to as an equilibrium steady state. Unless otherwise stated, the word `equilibrium' appearing in this article will be always understood as
an equilibrium steady state. However, if detailed balance is violated, then the enzyme system depicted in Fig.~\ref{mechanism} is an open system that approaches a non-equilibrium steady state (NESS) \cite{Hill, Wyman}. This is the scenario in enzyme kinetics \cite{Fersht, Segel}.

\subsection{Steady state rate formula}
\begin{figure}[ht]
\begin{center}
\centerline{\includegraphics[width=.5\textwidth]{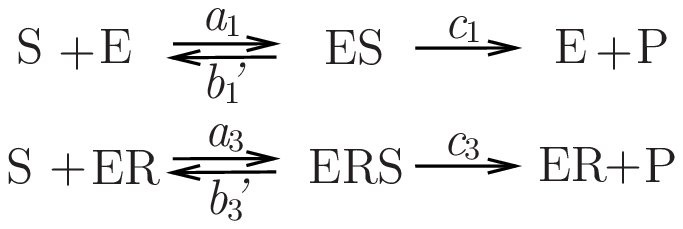}}
\caption{}\label{reduction}
\end{center}
\end{figure}

In the two extreme cases when the modifier concentration approaches zero or infinity, the general modifier mechanism reduces to the single-substrate single-product Michaelis-Menten mechanism depicted in Fig.~\ref{reduction}. The steady state product rate can be easily calculated in such cases. In fact, when the modifier concentration $[R]$ is zero, the steady state rate is
\begin{equation}\label{v0}
v_0 = c_1\mu_{ES} = \frac{c_1a_1[S]}{a_1[S]+b_1}.
\end{equation}
Similarly, when the modifier concentration $[R]$ is very large, the steady state rate is
\begin{equation}\label{vinfty}
v_\infty = c_3\mu_{ERS} = \frac{c_3a_3[S]}{a_3[S]+b_3}.
\end{equation}
The steady state rates $v_0$ and $v_\infty$ are collectively referred to as limiting product rates hereinafter.

Let $\nu=\mu_{ER}+\mu_{ERS}$ be the sum of the steady state probabilities of enzyme states $ER$ and $ERS$. It follows from the definition of product rate \eqref{velocity}, the circulation equation \eqref{circulation}, and the expressions of two limiting product rates \eqref{v0} and \eqref{vinfty} that
\begin{equation}\label{deltav_NESS}
v - v_0 = (v_\infty - v_0)\nu + (\frac{v_\infty}{a_3}-\frac{v_0}{a_1})\frac{J}{[S]}.
\end{equation}
The quantity
\begin{equation}
\Delta v = v-v_0
\end{equation}
in the left-hand side of Eq.~\eqref{deltav_NESS} is of special significance, since the sign of $\Delta v$ reflects the role of the modifier. A positive sign of $\Delta v$ indicates that the modifier is an activator, while a negative sign of $\Delta v$ indicates that the modifier is an inhibitor.

It is important to notice that the steady state rate formula \eqref{deltav_NESS} has clear biophysical significance. It is easy to see from this expression that the role of the modifier is determined by the competition of the first term of the right-hand side which does not include the net flux $J$ and the second term which includes the net flux $J$.

\section{Kinetic analysis}

\subsection{Kinetic analysis in an equilibrium steady state}
In an equilibrium steady state, the catalytic cycle in Fig.~\ref{cycle} is in detailed balance and there is no net flux. In this case, the steady state rate formula \eqref{deltav_NESS} suggests that
\begin{equation}
v = v_0+(v_\infty - v_0)\nu.
\end{equation}
Based on the condition of detailed balance \eqref{DB}, we have the following equation in which the dependence of the steady state rate $v$ on the modifier concentration $[R]$ becomes extremely clear:
\begin{equation}
v = \frac{v_\infty[R]+v_0K}{[R]+K},
\end{equation}
where $K=(b_1b_2a_3+a_1b_2a_3[S])/(a_1a_2b_3+a_1a_2a_3[S])$ is a constant if the substrate concentration $[S]$ is fixed.

\begin{figure}[ht]
\begin{center}
\centerline{\includegraphics[width=.6\textwidth]{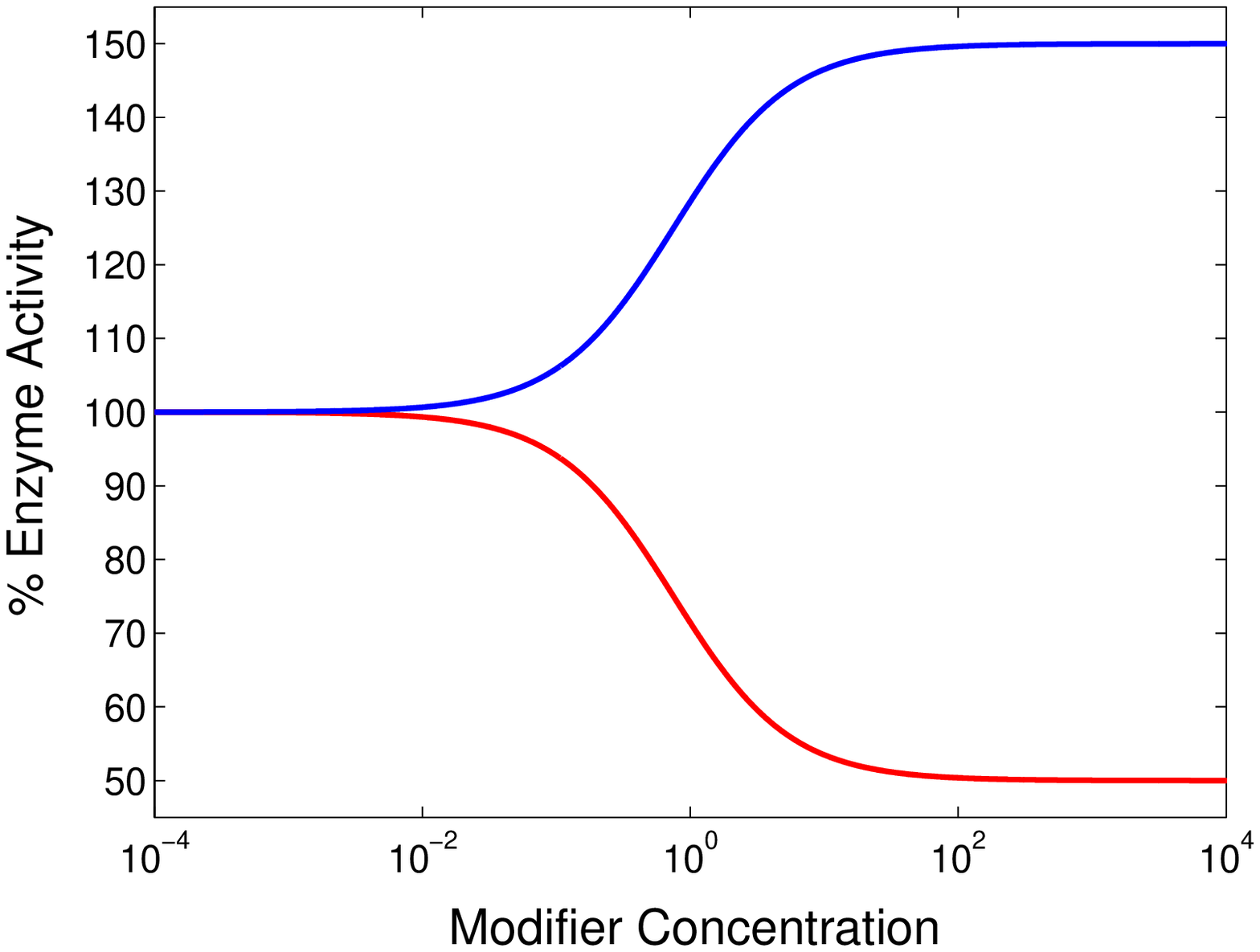}}
\caption{}\label{equilibrium}
\end{center}
\end{figure}

Notice that if $[S]$ is fixed, the product rate in equilibrium exhibits a hyperbolic dependence on $[R]$. In this case, the modifier acts as an overall activator or inhibitor for all its possible concentration values, depending on whether $v_0$ is smaller or greater than $v_\infty$. In other words, when the reaction system is in an equilibrium steady state, to be an activator or an inhibitor is an intrinsic property of the modifier. Fig.~\ref{equilibrium} is an illustration of the above conclusion.

\subsection{Kinetic analysis in NESS}\label{kineticsNESS}
In a non-equilibrium steady state, the net flux $J$ will no longer vanish. As is mentioned above, the steady state rate formula \eqref{deltav_NESS} has clear biophysical significance and the role of the modifier is determined by the competition of the equilibrium term
\begin{equation}
v_{eq} = (v_\infty - v_0)\nu
\end{equation}
which reflects the contribution of the equilibrium kinetics and the non-equilibrium term
\begin{equation}
v_{neq} = (\frac{v_\infty}{a_3}-\frac{v_0}{a_1})\frac{J}{[S]}
\end{equation}
which reflects the contribution of the non-equilibrium kinetics due to the non-zero net flux $J$.

In the following discussion, for simplicity, we assume $v_\infty < v_0$. The other case of $v_\infty > v_0$ can be discussed in the same way. Under this assumption, the equilibrium term $v_{eq}$ is negative for all possible values of $[R]$. It is readily verified that the equilibrium term $v_{eq}$, which behaves like the product rate in an equilibrium steady state, exhibits an approximately hyperbolic dependence on $[R]$, as is illustrated by the red curve in Fig.~\ref{switching}. On the other hand, the expression of $J$ in the supplementary material indicates that the non-equilibrium term $v_{neq}$ exhibits a bell shape vanishing at both zero and infinity as a function of $[R]$, as is illustrated by the green curve in Fig.~\ref{switching}.

\begin{figure}[ht]
\begin{center}
\centerline{\includegraphics[width=.6\textwidth]{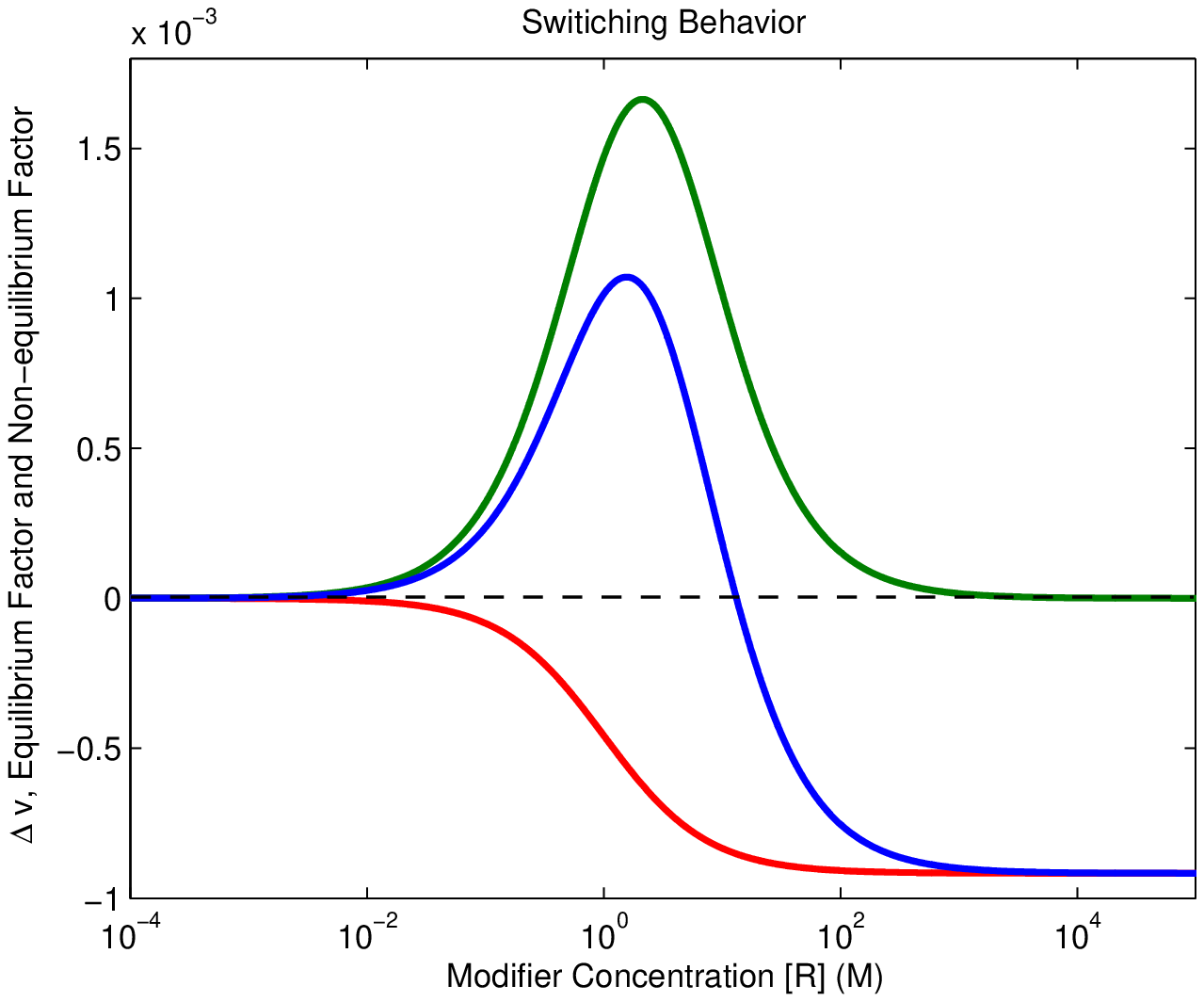}}
\caption{}\label{switching}
\end{center}
\end{figure}

Under certain conditions described in the next section (Sec.~\ref{swibehavior}), the non-equilibrium term $v_{neq}$ may be positive and play a dominating role compared to the negative contribution of the equilibrium term $v_{eq}$, as is illustrated in Fig.~\ref{switching}. In this case, $v-v_0$ is positive when $[R]$ is small and negative when $[R]$ is sufficiently large. The quantity $\Delta v$ will change the sign somewhere in the range of the modifier concentration $[R]$. In other words, the role played by the modifier will convert from an activator to an inhibitor, as is illustrated by the blue curve in Fig.~\ref{switching}.

Unexpectedly, the traditional point of view that being an activator or an inhibitor is an intrinsic property of the modifier does not hold in NESS. The role of the modifier depends greatly on the modifier concentration $[R]$. The new phenomenon in NESS described above, in which the role of the modifier converts between an activator and an inhibitor, is referred to as switching phenomenon throughout this paper. We will discuss the switching behavior of the modifier in much more detail in Sec.~\ref{swibehavior}.

Before leaving this section, we would like to point out that the switching behavior of the modifier may provide an explanation for the phenomenon of the side-effects caused by excessive intake of drugs. In fact, drugs are typical modifiers in pharmacology. If a drug operates in a mechanism discussed in this article that is far from equilibrium, the above analysis predicts that the role of the drug may convert when the drug concentration varies. Specifically, when a switching modifier is a drug designed to inhibit a particular enzyme in vitro, the concentration of the drug must be controlled within a certain range to maintain the inhibition effect. An exceptionally high concentration may cause the dangerous side-effects in which the inhibitor can produce an activation effect.

\section{Classification of the Modifier}

\subsection{Preparations}
We are now prepared to classify all possible steady state behavior of modifiers involved in the general modifier mechanism of Botts and Morales depicted in Fig.~\ref{mechanism}. We will give a quantitative criterion for various kinds of non-equilibrium kinetics in terms of the rate constants, $a_i$ and $b_i$, and the limiting product rates, $v_0$ and $v_\infty$. To make our discussion friendly to those unfamiliar with tedious mathematical tools, we would like to present the results here and put the mathematical treatment in the supplementary material.

The following three quantities play a key role in our classification:
\begin{equation}
\begin{split}
\Delta_1 &= v_\infty-v_0, \\
\Delta_2 &= q'(v_\infty-v_0) + k(\frac{v_\infty}{a_3}-\frac{v_0}{a_1}), \\
\Delta_3 &= (q-q')(v_\infty-v_0) - k(\frac{v_\infty}{a_3}-\frac{v_0}{a_1}),
\end{split}
\end{equation}
where $k = a_1a_2b_3b_4 - b_1b_2a_3a_4$ and $q,q^{\prime}\;(q^{\prime}<q)$ are positive parameters depending
on the rate constants and the substrate concentration $[S]$.

Virtually, in terms of the sign of the above three discriminants, we can classify the behavior of the modifier with completeness and clarity.

\subsection{Hyperbolic behavior: $\Delta_2\Delta_3>0$}
If $\Delta_2\Delta_3>0$, the steady state product rate exhibits an approximately hyperbolic dependence on $[R]$. In this case, the modifier behaves in the same way as it does in an equilibrium steady state except that the modification is stronger or weaker due to the contribution of the net flux $J$. The blue curves in Fig.~\ref{classification}(A) and Fig.~\ref{classification}(B) illustrate the hyperbolic behavior of the modifier as an overall activator and an overall inhibitor, respectively.

\subsection{Bell-shaped behavior: $\Delta_1\Delta_3<0$}
If $\Delta_1\Delta_3<0$, the steady state product rate exhibits a bell-shaped dependence on $[R]$. The bell-shaped behavior differs from the hyperbolic behavior in that the product rate will reach an extreme value which excels the limiting rate $v_\infty$ with the increase of $[R]$. The red curves in Fig.~\ref{classification}(A) and Fig.~\ref{classification}(B) illustrate the bell-shaped behavior of the modifier as an overall activator and an overall inhibitor, respectively.

Although both the hyperbolic-behaved and bell-shaped-behaved modifiers are overall activators or inhibitors for all possible values of $[R]$ when $[S]$ is fixed, there exist crucial differences between them. For the hyperbolic kinetics, an inordinately large increase of $[R]$ is necessary to bring about even a comparatively modest change in the enzyme activity. For the bell-shaped kinetics, on the contrary, a modest increase of $[R]$ will make the enzyme activity exceed its limit value $v_\infty$. Therefore, if the requirement for effective regulation (activation or inhibition) of the living body is needed, then bell-shaped modification will be a good choice.

\subsection{Switching behavior: $\Delta_1\Delta_2<0$}\label{swibehavior}
If $\Delta_1\Delta_2<0$, the quantity $\Delta v$ will change the sign somewhere in the range of the modifier concentration. In this case, the role played by the modifier will convert from an activator to an inhibitor or vice versa. Therefore in NESS, as is pointed out in Sec.~\ref{kineticsNESS}, the effect of activation or inhibition should not be viewed as an intrinsic property of the modifier.

\begin{figure}[ht]
\begin{center}
\centerline{\includegraphics[width=1.2\textwidth]{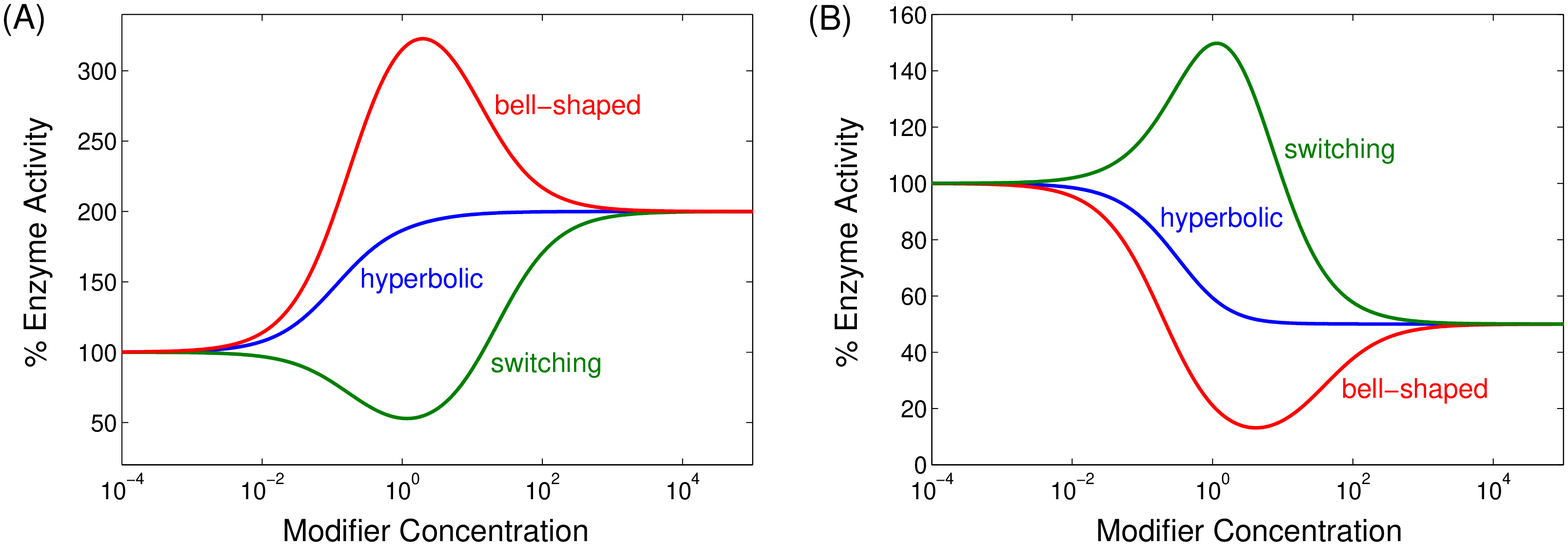}}
\caption{}\label{classification}
\end{center}
\end{figure}

The switching behavior of the modifier is illustrated by the green curves in Fig.~\ref{classification}(A) and Fig.~\ref{classification}(B). Fig.~\ref{classification}(A) illustrates the transition from an inhibitor to an activator, while Fig.~\ref{classification}(B) illustrates the transition from an activator to an inhibitor.

\subsection{Classification of the modifier in terms of $[S]$}
\begin{figure}[ht]
\begin{center}
\includegraphics[width=.6\textwidth]{{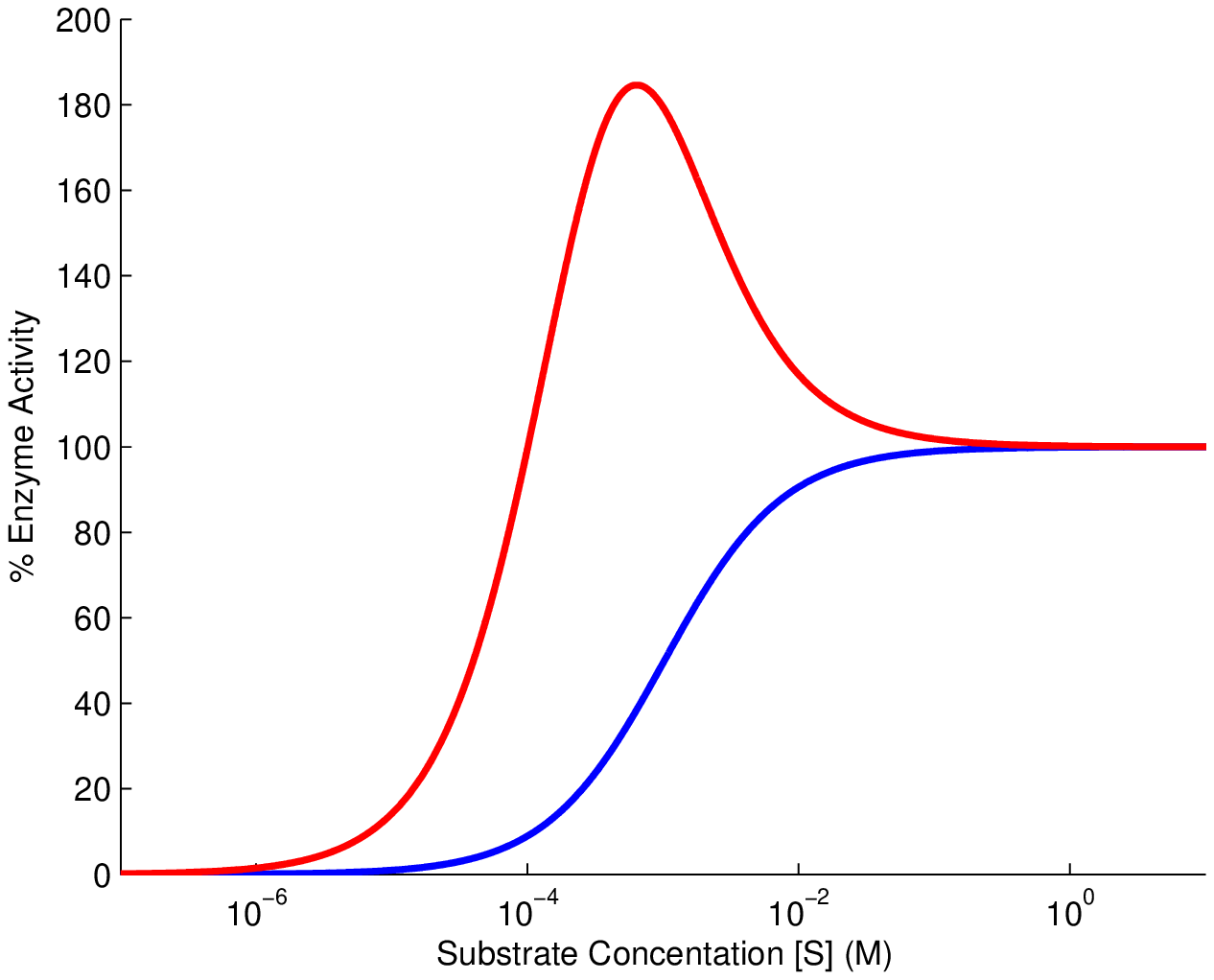}}
\caption{}\label{substrate}
\end{center}
\end{figure}

Incidentally, the behavior of the modifier can be classified according to the regulation of enzyme activity by the substrate concentration $[S]$. In fact, the steady state rate $v$ may exhibit an approximately hyperbolic or a bell-shaped dependence on $[S]$. The bell-shaped dependence differs from the hyperbolic one in that the steady state rate will reach an extreme value which excels the saturated rate with the increase of $[S]$. This conclusion is illustrated in Fig.~\ref{substrate}, where the blue and red curves represent the above two cases, respectively.

\section{Discussion}

\subsection{Conditions for hyperbolic behavior}
In the following discussion, the bell-shaped behavior and the switching behavior of the modifier are collectively referred to as non-hyperbolic behavior. As is analyzed above, non-hyperbolic behavior never appears in an equilibrium steady state. In this section, we will discuss another case in which non-hyperbolic behavior never occurs.

To this end, we rewrite the steady state rate formula \eqref{deltav_NESS} as
\begin{equation}\label{deltav2}
\Delta v = (v_\infty - v_0)(\nu + \frac{J}{a_3[S]}) + \frac{v_0}{[S]} (\frac{1}{a_3}-\frac{1}{a_1})J.
\end{equation}
It is readily verified that the first term on the right-hand side of this equation exhibits a hyperbolic dependence on $[R]$. Hence non-hyperbolic behavior will never occur as long as the second term vanishes.

In an equilibrium steady state, the second term vanishes due to the zero net flux. Another crucial situation where the second term will vanish arises when $a_1=a_3$, which implies that whether the modifier is bound to the enzyme or not will not affect the binding affinity of the substrate binding site. The above argument leads to the fact that when the binding rate of the substrate is irrelevant of the modifier, only hyperbolic behavior will occur, even if the enzyme system is in NESS.

\subsection{Examples of switching behavior}
As is pointed out in the last paragraph of Sec.~\ref{kineticsNESS}, drugs are typical modifiers in pharmacology. If a drug operates in a mechanism discussed in this article that is far from equilibrium, the above analysis predicts that the role of the drug may convert when the drug concentration varies.

ATPase activity associated with P-glycoprotein (Pgp) is characterized by three drug-dependent phases: basal (no drug), drug-activated, and drug-inhibited. The communication between the drug-binding site and the ATP hydrolytic site on a Pgp enzyme makes the reaction system a general modifier mechanism of Botts and Morales where ATP acts as the substrate and the drug acts as the modifier \cite{Al-Shawi1988, Al-Shawi2003}.

\begin{figure}[ht]
\begin{center}
\centerline{\includegraphics[width=.6\textwidth]{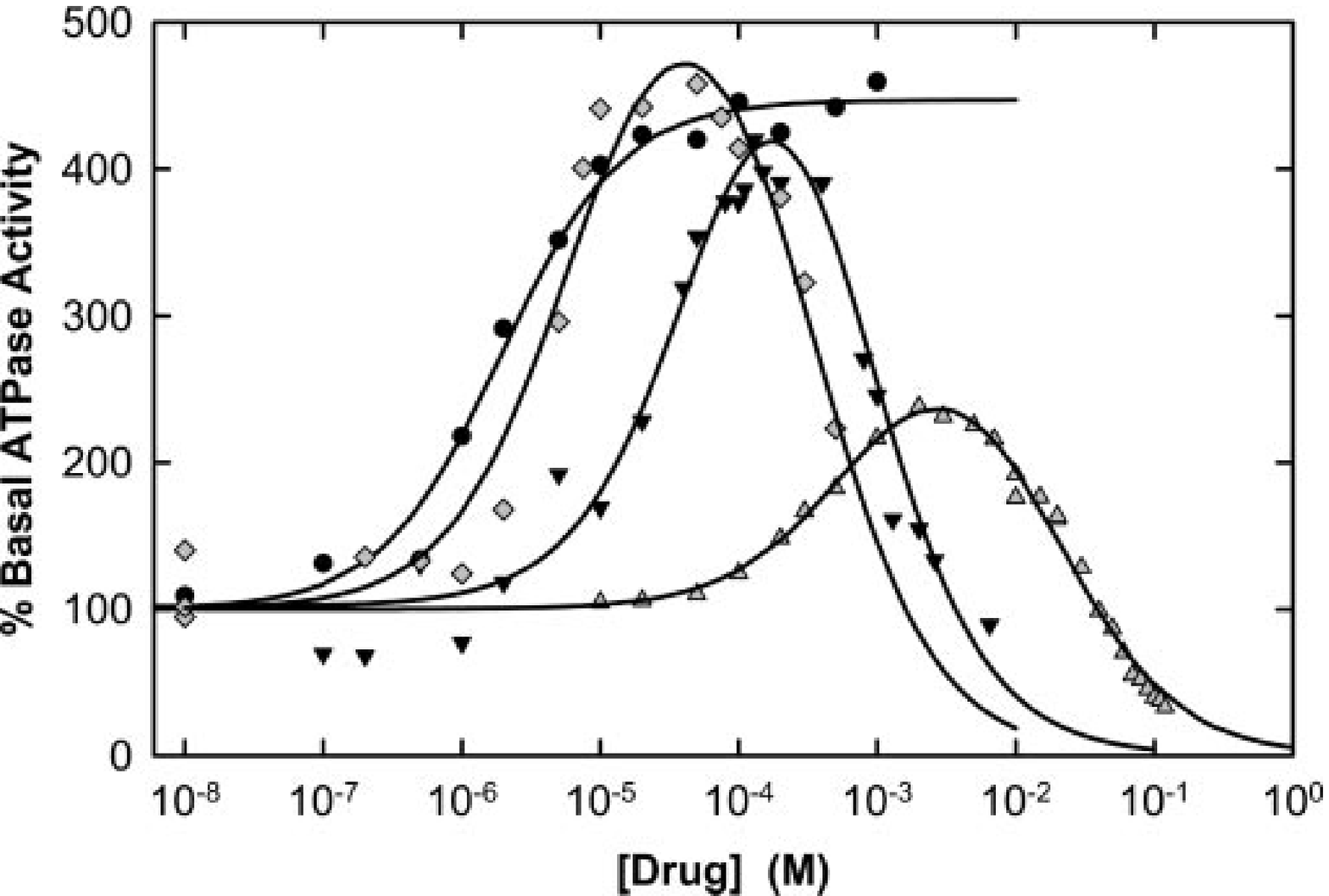}}
\caption{}\label{drug}
\end{center}
\end{figure}

Experimental data show that both hyperbolic activation and switching phenomena occur under the experimental condition of pH 7.4 and 37$^\circ$C. The four curves in Fig.~\ref{drug}, which was first generated by Al-Shawi et al. \cite{Al-Shawi2002, Al-Shawi2003}, represent the variation trends of enzyme activity versus concentrations of different types of drugs, namely valinomycin, verapamil, SL-verapamil, and colchicine, respectively. The occurrence of switching phenomenon in this experiment strongly supports the analysis of non-equilibrium kinetics of the general modifier mechanism presented in this paper.

\subsection{Applications}
\begin{figure}[ht]
\begin{center}
\centerline{\includegraphics[width=.5\textwidth]{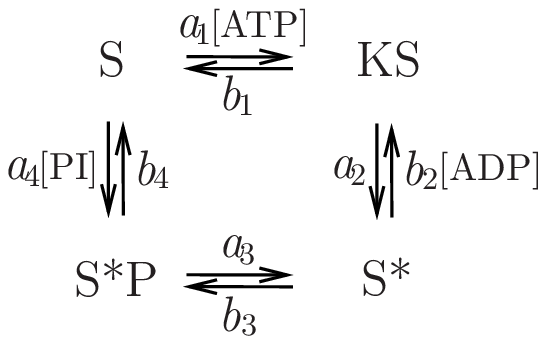}}
\caption{}\label{phosphorylation}
\end{center}
\end{figure}

We have seen that the core of the general modifier mechanism of Botts and Morales is the four-state cyclic reaction system depicted in Fig.~\ref{cycle}. Actually, the four-state cyclic topology is fundamental, since it models almost all possible reaction mechanism of proteins with two binding sides. Particularly, in living cells, biochemical processes that involve a four-state cyclic reaction system are very common. For example, the fundamental phosphorylation-dephosphorylation cycle illustrated in Fig.~\ref{phosphorylation} constitutes a four-state cyclic reaction system \cite{Qianbook}.

\begin{figure}[ht]
\begin{center}
\centerline{\includegraphics[width=.5\textwidth]{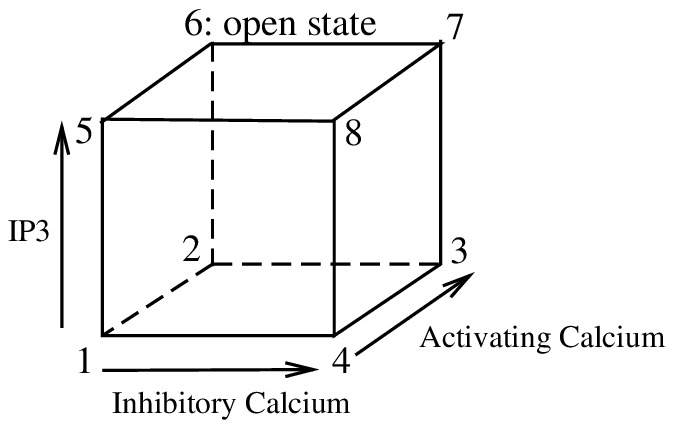}}
\caption{}\label{mymodel}
\end{center}
\end{figure}

Furthermore, in many biochemical reaction systems, proteins possess three or more binding sides. An eight-state cubic reaction mechanism is proposed to describe such kind of systems. A typical example of the eight-state cubic topology is the subunit of the inositol 1,4,5-trisphosphate receptor (IP$_3$R) which is a channel located in the endoplasmic reticulum that releases Ca$^{2+}$ ions. Structurally, the IP$_3$R is a large homomeric tetramer of four subunits forming a single ion-conducting channel \cite{structure}. The gating of IP$_3$R channels requires that three or all of the four subunits are at the open state \cite{Shuai}. Each subunit has one binding site for IP$_3$ and two binding sites for Ca$^{2+}$. Thus there are eight possible states for the subunit, which is illustrated in Fig.~\ref{mymodel}. Binding with IP$_3$ `potentiates' the subunit. The two calcium binding sites activate and inactivate the subunit, and a subunit is activated when IP$_3$ and the activating calcium site are bound but the inactivating site is unbound \cite{DeYKeiz}.

Besides, a great number of proenzymes activation mechanisms have one or two, identical or different modifiers acting on the activating and the activated enzymes. It turns out that such activation mechanisms may be considered, really or formally, as particular cases of the general eight-state cubic reaction mechanism \cite{Varon2007}.

Mathematically, it can be proved by the theory of Markov chains that if the binding of one site is irrelevant to that of the other two sites, the eight-state cubic reaction system depicted in Fig.~\ref{mymodel} can be reduced to the `combination' of a uni-unimolecular reaction system and a four-state cyclic reaction system. Thus the four-state catalytic cycle discussed in this article plays a basic role in analyzing the eight-state cubic reaction systems.

\section{Conclusion}
% Introduce of the net flux
In our present work, we introduce the net flux $J$ and propose a method in dealing with the cyclic reaction systems. Early approaches on the general modifier mechanism \cite{Fontes, Laidler} always give so many unreadable expressions and complicated situations that the hidden biophysical meanings are extremely hard to clarify. In this paper, however, with the notion of the net flux, the expression of steady state product rate has clear biophysical significance. We would like to emphasize that if the reaction kinetics is not too complex, the approach adopted in this paper gains an advantage of being neat and concise over the traditional approaches, e.g., the King-Atman method \cite{Cornish-Bowden, Qianbook} and numerical computation.

% Significance of circulation approach
In addition, we would like to point out that the single-molecule enzyme kinetics is now attracting broad scientific attention \cite{Nobel}. The general discussion in this article is based on the four-state catalytic cycle depicted in Fig.~\ref{cycle}. What we concentrate on is the transition among various states of a single enzyme molecule. The mathematical theory of non-equilibrium Markov chains and the concept of net flux along a cycle render possible a deep understanding of a number of interesting phenomena that never appear in equilibrium systems, such as the switching behavior of the modifier involved in the general modifier mechanism of Botts and Morales. Surely, applications of the non-equilibrium theory of Markov chains are not limited to the model discussed in this paper. It is expected to become a fundamental and useful tool in studying the single-molecule enzyme systems involved in signal transduction, gene regulation and synthetic biology.

\section*{Acknowledgements}
The authors thank Professor Hong Qian and Min Qian for helpful discussions and thank Dr. Hao Ge for valuable suggestions. This work is partly supported by the NSFC 10701004 and NSFC 10871009.

%% Enter the largest bibliography number in the facing curly brackets
%% following \begin{thebibliography}

\section*{Figure captions}
Fig.~\ref{mechanism}.
General modifier mechanism of Botts and Morales. The symbols $E$, $S$, $R$ and $P$ stand for the enzyme molecule, the substrate molecule, the modifier molecule and the product molecule, respectively, while the composite symbols $ES$, $ER$ and $ERS$ represent the corresponding complexes. The symbols $a_i$, $b_i'$ and $c_i$ are rate constants.

Fig.~\ref{cycle}.
Catalytic cycle of the general modifier mechanism depicted in Fig.~\ref{mechanism}. Since both reaction rates $b_1'$ and $c_1$ ($b_3'$ and $c_3$) in Fig.~\ref{mechanism} contribute to the transition from $ES$ ($ERS$) to $E$ ($ER$), we use the symbol $b_1$ ($b_3$) to represent $b_1'+c_1$ ($b_3'+c_3$).

Fig.~\ref{reduction}.
Two extreme cases of the general modifier mechanism depicted in Fig.~\ref{mechanism}. When the modifier concentration approaches zero or infinity, the general modifier mechanism reduces to the single-substrate single-product Michaelis-Menten mechanism depicted in this figure.

Fig.~\ref{equilibrium}.
Product rate (enzyme activity) in equilibrium versus modifier concentration. The initial enzyme activity is scaled to 1=100\%. In an equilibrium steady state, the modifier acts as an overall activator ($v_\infty>v_0$) or inhibitor ($v_\infty<v_0$) for all possible values of $[R]$, as is shown by the blue and red curves, respectively. The rate constants are chosen as $a_1=1, a_2=2, b_3=2, b_4=1, b_1=2, b_2=1, a_3=2, a_4=1, c_1=1.5, c_3=1.5, [S]=1$ for the case of activation. The rate constants are chosen as $a_1=1, a_2=2, b_3=2, b_4=1, b_1=2, b_2=1, a_3=2, a_4=1, c_1=1.5, c_3=0.5, [S]=1$ for the case of inhibition.

Fig.~\ref{switching}.
The switching behavior of the modifier in NESS when $v_\infty < v_0$. The red and green curves represent the equilibrium and non-equilibrium terms respectively and the blue curve represents the quantity $\Delta v = v - v_0$. The rate constants are chosen as $a_1=1, a_2=1, b_3=1, b_4=1, b_1=0.2, b_2=0.2, a_3=0.1, a_4=0.5, c_1=1, c_3=4.5833, [S]=1$ for this graph.

Fig.~\ref{classification}.
Enzyme activity in NESS versus modifier concentration.
(A) Enzyme activity in NESS versus modifier concentration when $v_\infty > v_0$. The initial enzyme activity is scaled to 1=100\%.
The rate constants are chosen as $a_1=20, a_2=20, b_3=10, b_4=10, b_1=1, b_2=1, a_3=50, a_4=10, c_1=1, c_3=2.3, [S]=1$ for the hyperbolic behavior.
The rate constants are chosen as $a_1=1, a_2=10, b_3=20, b_4=30, b_1=1, b_2=1, a_3=10, a_4=1, c_1=1, c_3=21, [S]=1$ for the bell-shaped behavior.
The rate constants are chosen as $a_1=1, a_2=10, b_3=10, b_4=10, b_1=1, b_2=1, a_3=0.5, a_4=1, c_1=1, c_3=3, [S]=1$ for the switching behavior.
(B) Enzyme activity in NESS versus modifier concentration when $v_\infty < v_0$.
The rate constants are chosen as $a_1=10, a_2=10, b_3=10, b_4=10, b_1=1, b_2=1, a_3=10, a_4=10, c_1=1, c_3=0.9, [S]=1$ for the hyperbolic behavior.
The rate constants are chosen as $a_1=1, a_2=10, b_3=50, b_4=50, b_1=1, b_2=1, a_3=20, a_4=1, c_1=1, c_3=0.9, [S]=1$ for the bell-shaped behavior.
The rate constants are chosen as $a_1=1, a_2=1, b_3=1, b_4=1, b_1=0.2, b_2=0.2, a_3=0.1, a_4=0.5, c_1=1, c_3=4.6, [S]=1$ for the switching behavior.

Fig.~\ref{substrate}.
Product rate (enzyme activity) in NESS versus substrate concentration. The saturated enzyme activity is scaled to 1=100\%. The blue and red curves represent the hyperbolic and bell-shaped dependence, respectively. The rate constants are chosen as $a_1=1, a_2=1, b_3=1, b_4=1, b_1=1, b_2=1, a_3=1, a_4=2, c_1=390, c_3=10, [R]=1$ for the hyperbolic dependence. The rate constants are chosen as $a_1=1, a_2=1, b_3=1, b_4=1, b_1=0.2, b_2=0.2, a_3=0.1, a_4=0.2, c_1=1150, c_3=10, [R]=1$ for the bell-shaped dependence.

Fig.~\ref{drug}.
Effects of drugs on Pgp ATPase activity. The ATPase activities of MDR1 reconstituted into mixed lipid vesicles were measured as a function of added drug concentration at pH 7.4 and 37$^\circ$ C \cite{Al-Shawi2002, Al-Shawi2003}. \emph{black circles}, valinomycin; \emph{inverted black triangles}, verapamil; \emph{gray diamonds}, SL-verapamil; \emph{gray triangles}, colchicine. Data for verapamil, SL-verapamil, and colchicine are from Omote and Al-Shawi \cite{Al-Shawi2002}.

Fig.~\ref{phosphorylation}.
A typical cellular biochemical switch consisting of a phosphorylation-dephosphorylation cycle. The substrate molecule $S$ may be a protein or other signaling molecule. If $S$ is a protein then the phosphorylation of $S$ is catalyzed by a protein kinase $K$ and the dephosphorylation is catalyzed by a protein phosphatase $P$. The entire cycle is accompanied by the reaction ATP $\rightleftharpoons$ ADP+PI.

Fig.~\ref{mymodel}.
Transition diagram for the eight-state subunit of IP$_3$ receptor channel. When the system is at state $6$ (one IP$_3$ and one activating Ca$^{2+}$ bound), the subunit is activated.

\end{document}